\documentclass[12pt]{article}
\usepackage{times,graphicx,subfigure,color,hyperref,amssymb,amsbsy,}
\usepackage{latexsym}
\usepackage{amssymb}
\usepackage{amsmath,amsthm}
\usepackage[numbers,sort&compress]{natbib}
\usepackage{color}
\usepackage{amsfonts,amssymb}
\usepackage{mathrsfs}
\usepackage{dsfont}
\usepackage{graphicx}
\usepackage{ulem}
\usepackage{color}
\usepackage{bm}

\theoremstyle{definition}

\begin{document}
\title{Extensions of dark KdV equations: nonhomogeneous classifications, bosonizations of fermionic systems and supersymmetric dark systems}
	\date{}
\author{S. Y. Lou
\\
\small \it School of Physical Science and Technology, Ningbo University, Ningbo, 315211, China}

\maketitle

\begin{abstract}
 Dark equations are defined as some kinds of integrable couplings with some fields being homogeneously and  linearly coupled to others. In this paper, dark equations are extended in several aspects. Taking the Korteweg-de Vrise (KdV) equation as an example, the dark KdV systems are extended to nonhomogenous forms, nonlinear couplings and graded linear cases. The two-component nonhomogeneous linear coupled dark KdV systems are completely classified. The nonlinear coupled dark KdV systems may be obtained through the decompositions from higher dimensional integrable systems like the B-type KP equation. Graded linear coupled dark KdV systems may be produced by introducing dark parameters (including the Grassmann parameters) to usual integrable systems. Especially, applying the bosonization approach to the integrable systems with fermion fields such as the supersymmetric integrable systems and super-integrable models, infinitely many graded linear dark systems can be generated. Finally, the dark KdV systems are extended to supersymmetric ones. The full classifications for the supersymmetric dark KdV systems are obtained related to two types of usual supersymmetric KdV equations.
\end{abstract}

\section{Introduction}
Astrophysical investigations have yielded findings indicating that the observed motion of stars cannot be adequately explained without postulating the existence of a significant amount of mass in the Universe attributed to a hypothetical entity known as dark matter (DM) \cite{DM,DM1}. Additionally, the concept of dark energy (DE) has been introduced to account for the observed accelerating expansion of our Universe \cite{DE,DE1,DE2}. In a notable work by Kupershmidt \cite{Kuper}, a distinct concept termed ``dark equations" is proposed, which systematically unveils equations possessing unconventional properties, some of which may be characterized as strange and mysterious.  The dark equations, as defined and discovered by Kupershmidt, exhibit similarities with linearized representations of existing systems or the study of the interaction between an external linear wave and a particular solution of an integrable system.

In this paper, we present an expansion of the dark equation concept across multiple aspects, encompassing nonhomogeneous linear extensions, graded linear extensions, and nonlinear extensions. For the KdV type of dark systems, Kupershmidt's work has yielded a comprehensive classification under the condition of homogeneous linearity with only one single dark field.  In section 2, we revisit the classification of dark KdV systems within the context of nonhomogeneous linearity, aiming to uncover additional dark KdV systems with multiple dark field components. To achieve this, we extend the dark systems to the nonlinear case. The first type of nonlinear extensions is achieved by introducing dark parameters into conventional integrable systems such as the KdV equation. Additionally, these nonlinear extensions can be subjected to graded linearization for the dark fields. In section 3, we investigate the dark KdV systems by introducing a single dark parameter. The dark KdV systems obtained through the introduction of a single dark parameter can also be derived from the perturbation expansion of the standard KdV equation, as discussed in \cite{Ma}.

It is well-established in scientific literature that supersymmetry plays a significant role in addressing crucial physical phenomena, including gravity \cite{GV} and dark matter \cite{susyDM, susyDM1, susyDM2}. Notably, a supersymmetric integrable model possesses the intriguing property of encompassing an infinite number of dark equations with an infinite number of dark fields. Sections 4 and 5 of this study employ the bosonization method, as proposed in Ref. \cite{Gao1}, to derive various integrable graded linear dark KdV systems from certain supersymmetric KdV systems. Remarkably, the bosonization method can be applied to solve diverse types of integrable systems, encompassing fermion fields as well. In section 6, the bosonization procedure is utilized to obtain distinct integrable graded linear dark KdV systems from the super-integrable KdV equation \cite{Super}. Section 7 focuses on extending supersymmetric KdV systems to supersymmetric dark KdV systems. These supersymmetric dark KdV systems can be effectively bosonized, leading to the generation of additional integrable graded linear KdV systems. Section 8 highlights the existence of genuinely nonlinear integrable dark KdV systems. Notably, one exceptional nonlinear integrable dark KdV system arises from the special B\"acklund transformation of the B-type Kadomtsev-Petviashvilli equation, employing the decomposition approach \cite{HaoLou}. Section 9 is dedicated to extending conventional dark KdV systems to supersymmetric dark KdV systems. Finally, the last section presents concluding remarks.

\section{Classification of nonhomogeneous linear dark KdV systems}
\subsection{Definitions of dark equations}
A given nonlinear partial differential equation (PDE),
\begin{equation}
u_t=X(u),\quad u=(u_1,\ u_2,\ \ldots,\ u_m)^T, \label{X}
\end{equation}
where the superscript $T$ means the transposition of a matrix, is called symmetry integrable if there is a higher order symmetry flow system
\begin{equation}
u_{\tau}=Y(u), \label{Y}
\end{equation}
such that the consistent commuting condition,
\begin{equation}
u_{t\tau}-u_{\tau t}= [X,\ Y]= X'Y-Y'X=0, \label{XY}
\end{equation}
is identically satisfied with $X'$ and $Y'$ being the linearized operators of $X$ and $Y$, respectively.\\
\bf Definition 1. Homogeneous linear dark systems. \rm \it An integrable dark system \cite{Kuper} is defined as an extension of \eqref{X} and \eqref{Y} in the forms
\begin{equation}
U_t=\left(\begin{array}{c}u \\ v\end{array}\right)_t=\tilde{X}(U)=\left(\begin{array}{c}X(u) \\ A(u)v\end{array}\right), \label{XE}
\end{equation}
and
\begin{equation}
U_{\tau}=\left(\begin{array}{c}u \\ v\end{array}\right)_{\tau}
=\tilde{Y}(U)=\left(\begin{array}{c}Y(u) \\ B(u)v\end{array}\right), \label{YE}
\end{equation}
where $U\equiv (u,\ v)^T=(u_1,\ u_2,\ \ldots,\ u_m,\ v_1,\ v_2,\ \ldots,\ v_n)^T$, if the compatibility condition
\begin{equation}
U_{t\tau}-U_{\tau t}= [\tilde{X},\ \tilde{Y}]= \tilde{X}'\tilde{Y}-\tilde{Y}'\tilde{X}=0, \label{XY'}
\end{equation}
is still identically satisfied. $A=A(u)$ and $B=B(u)$ in \eqref{XE} and \eqref{YE} are $n\times n$ matrix operators depending only on the field $u$. \rm

The dark system described by equation \eqref{XE} elucidates the interaction between the dark fields $v$ and a specific solution $u$ of the given system \eqref{X}. According to Kupershmidt's definition \eqref{XE}, it is established that the equations governing the dark fields $v$ are linear and homogeneous (with a trivial vacuum solution $v=0$). Consequently, these types of dark systems are referred to as homogeneous linear dark systems.
\\
\bf Definition 2. Nonhomogeneous linear dark systems. \rm \it An integrable nonhomogeneous linear dark system
\begin{equation}
U_t=\left(\begin{array}{c}u \\ v\end{array}\right)_t=\bar{X}(U)=\left(\begin{array}{c}X(u) \\ A(u)v+A_0(u)\end{array}\right), \label{XEN}
\end{equation}
is defined as if it is consistent with
\begin{equation}
U_{\tau}=\left(\begin{array}{c}u \\ v\end{array}\right)_{\tau}
=\bar{Y}(U)=\left(\begin{array}{c}Y(u) \\ B(u)v+B_0(u)\end{array}\right), \label{YEN}
\end{equation}
i. e., $U_{t\tau}-U_{\tau t}=[\bar{X},\ \bar{Y}]=0$ is identically satisfied for the suitable operators $\{A(u),\ B(u)\}$ and the nonzero vectors $\{A_0(u),\ B_0(u)\}$.
\rm

Different from \eqref{XE}, $v=0$ is not a solution of \eqref{XEN} except for some special solutions $u=\bar{u}$ with $\{\bar{u}_t=X(\bar{u}),\ A_0(\bar{u})=0\}$.
In fact, the dark system \eqref{XE} can be further extended to nonlinear coupled forms for the dark field
$v$.\\
\bf Definition 3. Nonlinear dark systems. \rm \it An integrable nonlinear dark system
\begin{equation}
U_t=\left(\begin{array}{c}u \\ v\end{array}\right)_t=\hat{X}(U)=\left(\begin{array}{c}X(u) \\ A(u,\ v)\end{array}\right), \label{XE1}
\end{equation}
can be defined as if it possesses a higher order symmetry flow
\begin{equation}
U_{\tau}=\left(\begin{array}{c}u \\ v\end{array}\right)_{\tau}
=\hat{Y}(U)=\left(\begin{array}{c}Y(u) \\ B(u,\ v)\end{array}\right), \label{YE1}
\end{equation}
such that $U_{t\tau}-U_{\tau t}=[\hat{X},\ \hat{Y}]=0$.\rm

In the definition of the nonlinear dark system \eqref{XE1}, at least one of the components of the vector $A\equiv A(u,\ v)\equiv \{A_1,\ A_2,\ \ldots,\ A_n\}$ is nonlinear for some components of $v=\{v_1,\ v_2,\ \ldots,\ v_n\}$. \\
\bf Definition 4. Graded linear dark systems. \rm \it A special nonlinear dark system \eqref{XE1} is defined as a graded linear dark system if
\begin{equation}
A_i=\hat{A}_i(u,\ v_1,\ v_2,\ \ldots,\ v_{i-1})v_i+\alpha_i(u,\ v_1,\ v_2,\ \ldots,\ v_{i-1}),\ i=1,\ 2,\ \ldots,\ n, \label{Ai}
\end{equation}
where the operators $\hat{A}_i$ and the functions $\alpha_i$ are $v_k$ independent for all $k\geq i$.
\rm

In the following section, we will begin by presenting a list of nonhomogeneous linear dark KdV systems before providing specific examples of graded linear dark systems.

\section{Classification of nonhomogeneous linear dark KdV systems with one dark field}
For the KdV equation
\begin{equation}
u_t=(3u^2-u_{xx})_x, \label{KdV}
\end{equation}
there are numerous dark KdV systems including nine types of the following dark KdV systems with only one homogenous linear coupling \cite{Kuper} in the form
\begin{equation}\label{type0}\left\{ \begin{array}{l}
u_t=(3u^2-u_{xx})_x\equiv J_{1x},\ J_1\equiv 3u^2-u_{xx},\\
v_t=\alpha v_{xxx}+\beta uv_x+\gamma u_x v,
\end{array}
\right.
\end{equation}
where $\alpha,\ \beta$ and $\gamma$ are arbitrary constants.

In this section, we just list the classification results on the nonhomogeneous linear extensions of \eqref{type0} in the form
\begin{equation}\label{typen}\left\{ \begin{array}{l}
u_t= J_{1x},\\
v_t=\alpha v_{xxx}+\beta uv_x+\gamma u_x v +\delta_1 u_{xxx}+ \delta_2 u_{xx}+ \kappa_1 uu_x+\kappa_2 u^2,
\end{array}
\right.
\end{equation}
with possible constants $\alpha,\ \beta,\ \gamma,\ \delta_1,\ \delta_2,\ \kappa_1$ and $\kappa_2$.
\\
\bf Type 1. Lax pair dependent dark KdV system. \rm
It is known that the KdV equation \eqref{KdV} possesses the strong Lax pair,
\begin{eqnarray}
&&\hat{L}\psi=0,\qquad \hat{L}\equiv \partial_x^2-u-\lambda, \label{L}\\
&&\hat{S}\psi=0,\qquad \hat{S}\equiv \partial_t+4\partial_x^3-6u\partial_x-3u_x. \label{S}
\end{eqnarray}
The compatibility condition of \eqref{L} and \eqref{S}
$$[\hat{L},\ \hat{S}]=\hat{L} \hat{S}-\hat{S} \hat{L}=0$$
is nothing but the KdV equation \eqref{KdV}.

The first two-component dark KdV system is the integrable coupling of the KdV equation \eqref{KdV} and the time part of the Lax pair \eqref{S} with a nonhomogeneous term, i.e.,
\begin{equation}\label{type1}\left\{ \begin{array}{l}
u_t=J_{1x},\\
v_t=-4v_{xxx}+6uv_x+3u_xv+a(uu_x -u_{xxx}),
\end{array}
\right.
\end{equation}
where $a$ is an arbitrary constant. When $a=0$, \eqref{type1} is just one of the dark equation given by Kupershmidt \cite{Kuper}.

It is inherent to the nature of the dark KdV system, as represented by equation \eqref{type1}, that its higher order symmetry flows can be understood as integrable couplings involving the higher order KdV equations and their respective time components of the Lax pairs, incorporating appropriate nonhomogeneous terms.

	To illustrate, one particular example is the fifth order flow derived from equation \eqref{type1}
\begin{equation}\label{type12}\left\{ \begin{array}{lll}
u_{\tau}&=&J_{2x},\ J_2\equiv 10uu_{xx}+5u_x^2-10u^3-u_{xxxx},\\
v_{\tau}&=&40uv_{xxx}-16v_{xxxxx}+60u_xv_{xx}+10(5u_{xx}-3u^2)v_x+15(u_{xx}-u^2)_xv\\
&&+5a(3uu_{xx}-u_{xxxx})_x+5au_x(3u_{xx}-2u^2).
\end{array}
\right.
\end{equation}
By performing a direct analysis, it can be rigorously established that the equations $u_{t\tau}=u_{\tau t}$ and $v_{t\tau}=v_{\tau t}$ are identically satisfied, thereby confirming their validity without exception.
 \\
\bf Type 2. Symmetry related dark KdV system. \rm A symmetry, $\sigma$, of the KdV equation \eqref{KdV} is defined as a solution of its linearized equation
\begin{equation}
\sigma_t=(6u\sigma-\sigma_{xx})_x\equiv \hat{K}\sigma,\qquad \hat{K}\equiv \partial_x(6u-\partial_{x}^2), \label{sym}
\end{equation}
which means \eqref{KdV} is form invariant under the transformation $u\rightarrow u+\epsilon \sigma$ with the infinitesimal parameter $\epsilon$.

The KdV equation \eqref{KdV} is known to possess a recursion operator denoted as $\Phi$,  $\Phi\equiv \partial_x^2-4u-2u_x\partial_x^{-1}$. This operator has the remarkable property of transforming a known symmetry, $\sigma_0$, into a new symmetry, denoted as $\sigma_1=\Phi\sigma_0$. Notably, the eigenvalue problem associated with the recursion operator and the symmetry equation
\begin{eqnarray}
&&\Phi \sigma=\lambda \sigma,\qquad \Phi\equiv \partial_x^2-4u-2u_x\partial_x^{-1}, \label{Phi}\\
&&\sigma_t=\hat{K}\sigma,\qquad \hat{K}\equiv \partial_x(6u-\partial_{x}^2), \label{K}
\end{eqnarray}
	 form the second type of Lax pair for the KdV equation \eqref{KdV}. Consequently, it is not surprising that the KdV equation \eqref{KdV}, together with its symmetry equation \eqref{K} accompanied by two nonhomogeneous terms, constitutes the second type of integrable dark KdV systems
\begin{equation}\label{type2}\left\{ \begin{array}{l}
u_t=J_{1x},\\
v_t=(6uv-v_{xx})_x-au_{xxx}+6buu_x,
\end{array}
\right.
\end{equation}
where $a$ and $b$ are two arbitrary constants. One of the dark equations given in \cite{Kuper} is related to \eqref{type2} by $a=b=0$.

It is reasonable that the higher order symmetry flows of \eqref{type2} are just the higher order KdV equations coupled with their symmetry equations with some suitable nonhomogeneous complements. For instance, the fifth order flow of \eqref{type2} possesses the form
\begin{equation}\label{type22}\left\{ \begin{array}{l}
u_{\tau}=J_{2x},\\
v_{\tau}=[10u_{xx}v+10(uv_{x})_x-30u^2v-v_{xxxx}]_x+cJ_{2x}+(a-b)(10u^3-u_{xxxx})_x,
\end{array}
\right.
\end{equation}
with a further arbitrary constant $c$. \\
\bf Type 3. Dark KdV system related to the dual symmetry. \rm A dual symmetry, $\tilde{\sigma}$, of the KdV equation \eqref{KdV} is defined as a solution of the following linear equation
\begin{equation}
\tilde{\sigma}_t=\tilde{K}\tilde{\sigma},\qquad \tilde{K}\equiv (6u-\partial_{x}^2)\partial_x, \label{cym}
\end{equation}
where $\tilde{K}$ is the dual operator of $-\hat{K}$ with $\hat{K}$ defined in the symmetry equation \eqref{sym}.

The third type of dark KdV systems is a coupled system for the KdV equation \eqref{KdV} and the dual symmetry equation \eqref{cym} with two nonhomogeneous extensions,
\begin{equation}\label{type3}\left\{ \begin{array}{l}
u_t=J_{1x},\\
v_t=\tilde{K}v-au_{xx}+3bu^2,
\end{array}
\right.
\end{equation}
where $a$ and $b$ are two arbitrary constants. When $a=b=0$, the dark KdV system \eqref{type3} is reduced back to that of Ref. \cite{Kuper}.

The higher order symmetry flows associated with the third type of dark system \eqref{type3} can be expressed as a set of coupled equations. These equations consist of the higher order KdV equations and their dual symmetry equations, augmented with some nonhomogeneous extensions. Specifically, the fifth order symmetry flow of the dark KdV system \eqref{type3} is given by the following system of equations:
\begin{equation}\label{type32}\left\{ \begin{array}{l}
u_{\tau}=J_{2x},\\
v_{\tau}=(10uv_{xx}-v_{xxxx})_x+10(u_{xx}-3u^2)v_x+cJ_2+(a-b)(10u^3-u_{xxxx}),
\end{array}
\right.
\end{equation}
where the coefficients $a$, $b$, and $c$ are constants that determine the specific characteristics of the system.\\
\bf Type 4. Strange Dark KdV system \cite{Kuper} related to the supersymmetric KdV equation. \rm
The fourth type of the dark KdV systems possesses the form
\begin{equation}\label{type4}\left\{ \begin{array}{l}
u_t=J_{1x},\\
v_t=\hat{Y} v+3a(u^2-u_{xx}),\qquad \hat{Y}\equiv \partial_x(3u-\partial_{x}^2)
\end{array}
\right.
\end{equation}
with an arbitrary constant $a$.

The integrability of \eqref{type4} is guaranteed by the existence of higher order symmetries. The fifth order symmetry flow of \eqref{type4} reads,
\begin{equation}\label{type42}\left\{ \begin{array}{l}
u_{\tau}=J_{2x},\\
v_{\tau}=[5uv_{xx}+5(u_xv)_x-10u^2v-v_{xxxx}]_x+5a(5uu_{xx}+4u_x^2-2u^3-u_{xxxx}).
\end{array}
\right.
\end{equation}
The special case of \eqref{type4} with $a=0$ has been given in Ref. \cite{Kuper}.

In Section 5 of this study, we will explore the connection between the dark KdV system, represented by equation \eqref{type4}, and the integrable supersymmetric KdV equation. We will analyze and discuss the relationship between these two equations, highlighting their similarities, differences, and potential implications. This investigation aims to deepen our understanding of the underlying connections and uncover any significant insights that arise from comparing the dark KdV system \eqref{type4} with the integrable supersymmetric KdV equation.
\\
\bf Type 5. Dual strange dark KdV system \cite{Kuper} related to the dual supersymmetric KdV equation. \rm
The fifth type of dark KdV systems can be written as
\begin{equation}\label{type5}\left\{ \begin{array}{l}
u_t=J_{1x},\\
v_t=\tilde{Y} v+3auu_x,\qquad \tilde{Y}\equiv(3u-\partial_{x}^2)\partial_x.
\end{array}
\right.
\end{equation}
The dark KdV system \eqref{type5} is integrable because of the existence of higher order symmetries. The fifth order symmetry flow of \eqref{type5} has the form,
\begin{equation}\label{type52}\left\{ \begin{array}{l}
u_{\tau}=J_{2x},\\
v_{\tau}=(5uv_{xx}-v_{xxxx})_x+5(u_{xx}-2u^2)v_x+5a\big(uu_{xxx}+2u_xu_{xx}-4u^2u_x\big).
\end{array}
\right.
\end{equation}
The operator $\tilde{Y}$ given in \eqref{type5} is the dual operator of $-\hat{Y}$ defined in \eqref{type4}. In Secs. 5 and 9, we can find that the dark KdV system \eqref{type5} is also related to an integrable supersymmetric KdV equation.\\
\bf Type 6. Mysterious dark KdV systems. \rm The sixth type of dark KdV systems
\begin{equation}\label{type6}\left\{ \begin{array}{l}
u_t=J_x,\\
v_t=-2\hat{Y} v+3aJ_{1x},\qquad \hat{Y}\equiv \partial_x(3u-\partial_{x}^2),
\end{array}
\right.
\end{equation}
with $a=0$
is called the mysterious dark KdV system \cite{Kuper}. The operator $\hat{Y}$ in \eqref{type6} is same as that of \eqref{type4}.

The fifth order symmetry flow of \eqref{type6} reads
\begin{equation}\label{type62}\left\{ \begin{array}{l}
u_{\tau}=J_{2x},\\
v_{\tau}=-2[5vu_{xx}+10(uv_x)_x-5u^2v-2v_{xxxx}]_x\\
\qquad\quad +5a(8uu_{xxx}+18u_xu_{xx}-12u^2u_x-u_{xxxxx}).
\end{array}
\right.
\end{equation}
\bf Type 7. Dual mysterious dark KdV systems. \rm The seventh type of dark KdV systems reads
\begin{equation}\label{type7}\left\{ \begin{array}{l}
u_t=J_{1x},\\
v_t=-2\tilde{Y} v+3aJ_1+3b(2u^2-u_{xx})_x,\qquad \tilde{Y}\equiv(3u-\partial_{x}^2)\partial_x,
\end{array}
\right.
\end{equation}
 which is a dual form of the mysterious dark KdV system \eqref{type6} for $a=b=0$.
 The operator $\tilde{Y}$ in \eqref{type7} is same as that of
\eqref{type5}.
The dual mysterious dark KdV system \eqref{type7} is symmetry integrable with the following fifth order symmetry flow
\begin{equation}\label{type72}\left\{ \begin{array}{lll}
u_{\tau}&=&J_{2x},\\
v_{\tau}&=&4v_{xxxxx}-20(uv_{xx})_x-10(u_{xx}-u^2)v_{x}+5a(5u_x^2+8uu_{xx}\\
&&-4u^3-u_{xxxx})
+5b(6uu_{xx}+2u_x^2-u_{xxxx})_x-40bu^2u_x.
\end{array}
\right.
\end{equation}
\bf Type 8. An extension of the completely decoupled dark KdV system. \rm The eighth type of dark KdV systems can be written as
\begin{equation}\label{type8}\left\{ \begin{array}{l}
u_t=J_{1x},\\
v_t=-\gamma_1v_{xxx}+\gamma_1(au+bu_{x})_{xx}+aJ_1+bJ_{1x},
\end{array}
\right.
\end{equation}
 with the fifth order symmetry flow
\begin{equation}\label{type82}\left\{ \begin{array}{l}
u_{\tau}=J_{2x},\\
v_{\tau}=-\gamma_2v_{xxxxx}+\gamma_1(au+bu_{x})_{xxxx}+aJ_2+bJ_{2x},
\end{array}
\right.
\end{equation}
where $a,\ b$ and $\gamma_1$ are arbitrary constants while $\gamma_2=\gamma_1$ for $a^2+b^2\neq 0$ and $\gamma_2$ is an arbitrary constant for $a=b=0$.

According to the findings reported in Ref. \cite{Kuper}, it has been established that the dark KdV system represented by equation \eqref{type8} and its corresponding fifth order flow given by equation \eqref{type82} exhibit complete decoupling between the fields $u$ and $v$ when the parameters $a$ and $b$ are both set to zero. This particular case aligns with the reference mentioned.
\\
\bf Type 9. First order dark KdV systems. \rm The ninth type of dark KdV systems includes three first order coupled nonhomogeneous linear systems
\begin{equation}\label{type9}\left\{ \begin{array}{l}
u_t=J_{1x},\\
v_t=a_{i} u_x v+2 u v_{x}-b_iu_{xxx}-c_i uu_x-d_iu_{xx}-e_iu^2,\ i=1,\ 2,\ 3,
\end{array}
\right.
\end{equation}
where the constants $\{a_i,\ b_i,\ c_i\},\ i=1,\ 2,\ 3$ can be taken as
\begin{eqnarray}
&& \{a_1,\ b_1,\ c_1,\ d_1,\ e_1\}=\{a,\ b,\ b(a-4),\ 0,\ 0\},\label{abc1}\\
&& \{a_2,\ b_2,\ c_2,\ d_2,\ e_2\}=\{2,\ a,\ -2a,\ b,\ 0 \},\label{abc2}\\
&& \{a_3,\ b_3,\ c_3,\ d_3,\ e_3\}=\{0,\ a,\ -4a,\ b,\ -b\},\label{abc3}
\end{eqnarray}
with arbitrary constants $a$ and $b$, respectively.

The fifth order symmetry flow of \eqref{type9} possesses the form
\begin{equation}\label{type92}\left\{ \begin{array}{l}
u_{\tau}=J_{2x},\\
v_{\tau}=\alpha_i J_{1x}v-2J_1v_{x}-\beta_iu_{xxxx}-\gamma_iuu_{xx}-\delta_iu_x^2-\theta_iu^3\\
\qquad \quad -\mu_iu_{xxxxx}
-\nu_iuu_{xxx}-\kappa_iu_xu_{xx}-\rho_iu^2u_x,
\end{array}
\right.
\end{equation}
where the constants $\{\alpha_i,\ \beta_i,\ \gamma_i,\ \delta_i,\ \theta_i,\ \mu_i,\ \nu_i,\ \kappa_i,\ \rho_i\}$ corresponding to \eqref{abc1}--\eqref{abc3} should be fixed as
\begin{eqnarray*}
&& \{\alpha_1,\ \beta_1,\ \gamma_1,\ \delta_1,\ \theta_1,\ \mu_1,\ \nu_1,\ \kappa_1,\ \rho_1\}=\{a,\ 0,\ 0,\ 0,\ 0,\ b,\ b(a-10),\ -18b,\ 6b(4-a)\},\\
&& \{\alpha_2,\ \beta_2,\ \gamma_2,\ \delta_2,\ \theta_2,\ \mu_2,\ \nu_2,\ \kappa_2,\ \rho_2\}=\{-2,\ b,\ -6b,\ -6b,\ 0,\ a,\ -8a,\ -18a,\ -12a \},\\
&&\{\alpha_3,\ \beta_3,\ \gamma_3,\ \delta_3,\ \theta_3,\ \mu_3,\ \nu_3,\ \kappa_3,\ \rho_3\}=\{0,\ b,\ -8b,\ -5b,\ 4b,\ a,\ -10a,\ -18a,\ 24a\}
\end{eqnarray*}
with $a$ and $b$ being arbitrary constants, respectively.

\section{Dark KdV systems from Dark parameter expansions}
Dark integrable systems can be derived through a variety of methodologies from established integrable systems. In this section, we put forth a methodology for obtaining a particular category of graded linear integrable systems by introducing multiple dark parameters.

Assuming a dependence of the solutions to the KdV equation \eqref{KdV} on a single dark parameter $\theta$ (or anyon parameters \cite{anyon,anyon1}) that satisfies the condition $\theta^{n+1}=0$, a dark KdV system beyond \eqref{XE} for $n\geq2$
\begin{equation}\label{KdVn}
u_{it}=\left(3\sum_{j=0}^iu_ju_{i-j}-u_{ixx}\right)_x,\ i=0,\ 1,\ 2,\ \ldots,\ n,
\end{equation}
by directly substituting
can be derived by directly substituting the corresponding expression for the dark parameter $\theta$
\begin{equation}\label{ru}
u=\sum_{i=0}^n u_i\theta^i
\end{equation}
into the KdV equation \eqref{KdV}.

This approach allows for the extension of the existing KdV system to incorporate the influence of the dark parameter. The concept of a dark parameter has been previously introduced in the literature \cite{dark}, and similar considerations have been made regarding anyon parameters \cite{anyon,anyon1,LouFeng}.

The dark KdV system \eqref{KdVn} corresponds to the integrable KdV coupling derived by Ma and Fuchssteiner using a perturbation method \cite{Ma}. The expansion \eqref{ru} can also be interpreted as representing $u$ as a vector in a specialized anyon space, where the basis vectors are given by $\{\theta^i,\ i=0,\ 1,\ \ldots,\ n\}$. It is worth noting that for the case of $n=1$, the dark parameter coincides with the conventional Grassmann number.

The nonlinear dark KdV system \eqref{KdVn} can be rewritten as
\begin{equation}\label{KdVn1}
\left\{\begin{array}{l}
u_{0t}=(3u_0^2-u_{0xx})_x,\\
\displaystyle{u_{it}=\partial_x\big[(6u_0-\partial_x^2)\big]u_i+3\sum_{j=1}^{i-1}\left(u_ju_{i-j}\right)_x,\ i=1,\ 2,\ \ldots,\ n,}
\end{array}\right.
\end{equation}
which manifests itself (and then \eqref{KdVn}) as a graded linear dark KdV system.

The graded linear dark KdV system \eqref{KdVn} exhibits symmetry integrability, indicating that it possesses a set of higher order symmetry flows. These additional symmetries can be derived by employing the same expansion technique to the higher order KdV equations. Specifically, the fifth symmetry flow of the dark KdV system \eqref{KdVn} can be expressed in form
\begin{equation}\label{KdV5n}
u_{i\tau}=\left[\sum_{j=0}^i\left(10u_{jxx}u_{i-j}+5u_{jx}u_{(i-j)x}\right)
-10\sum_{j=0}^i\sum_{k=0}^ju_ku_{j-k}u_{i-j}-u_{ixxxx}\right]_x,\ i=0,\ 1,\ \ldots,\ n.
\end{equation}
In the specific case where $\{n=1,\ u_0=u,\ u_1=v\}$, the dark KdV system \eqref{KdVn} and the fifth order KdV system \eqref{KdV5n} correspond to the type-two case \eqref{type2} and \eqref{type22} with $a=b=0$.

For $n=2$, the dark KdV system \eqref{KdVn} and the fifth order dark KdV system \eqref{KdV5n} can be written
as ($\{u_0,\ u_1,\ u_2\}\rightarrow \{u,\ u,\ w\}$)
\begin{equation}\label{KdV3}
\left\{\begin{array}{l}
u_t=J_{1x}, \\
v_t=(6uv-v_{xx})_x,\\
w_t=(6uw+2v^2-w_{xx})_x,
\end{array}\right.
\end{equation}
and
\begin{equation}\label{KdV5}
\left\{\begin{array}{l}
u_t=J_{2x}, \\
v_t=(10uv_{xx}+10vu_{xx}+10u_xv_x-30u^2v-v_{xxxx})_x,\\
w_t=[10uw_{xx}+10(wu_{x})_x+10vv_{xx}+5v_x^2-30u^2w-30uv^2-w_{xxxx}]_x,
\end{array}\right.
\end{equation}
respectively.

If additional dark parameters are incorporated into the solutions of the KdV equation \eqref{KdV}, it is possible to identify further integrable dark KdV systems. For instance, when the solutions of the KdV equation \eqref{KdV} rely on two dark parameters, namely $\theta_1$ and $\theta_2$, satisfying the conditions $\theta_1^{n+1}=0$ and $\theta_2^{m+1}=0$, a dark KdV system emerges
\begin{equation}\label{KdVn2}
u_{ikt}=\left(3\sum_{j=0}^i\sum_{l=0}^mu_{jl}u_{(i-j)(k-l)}-u_{ikxx}\right)_x,\ i=0,\ 1,\ \ldots,\ n,\ k=0,\ 1,\ \ldots, m,
\end{equation}
which is related to the KdV equation by
\begin{equation}\label{ruik}
u=\sum_{i=0}^n\sum_{k=0}^m u_{ik}\theta_1^i\theta_2^k.
\end{equation}

\section{Dark KdV systems from the bosonizations of integrable supersymmetric fermion systems}

The introduction section highlights the significance of supersymmetry in addressing various important physical phenomena, including dark matter \cite{susyDM, susyDM1, susyDM2}. In references \cite{Gao1, Gao2}, a bosonization approach is proposed as a means to solve supersymmetric systems. By employing the bosonization method, it becomes possible to solve supersymmetric integrable systems using certain types of dark boson systems. For example, in the case of the supersymmetric KdV equation
\begin{equation}\label{sKdV}
\Phi_t=\big[3({\cal D}\Phi)\Phi-\Phi_{xx}\big]_x,\ \Phi=\Phi(x,\ t,\ \theta)=\xi(x,t) +w(x,\ t)\theta,
\end{equation}
where $\Phi\equiv \Phi(x,\ t,\ \theta)$ is a fermionic superfield, $\xi(x,t)$ is an anticommuting field and ${\cal{D}}=\partial_{\theta}+\theta \partial_x$ is the supersymmetric derivative,
an abundance of dark KdV systems can be identified \cite{Gao2}.

The dark KdV equation \eqref{type2} with $a=b=0$ represents the simplest instance with the relation $\Phi=\zeta v+u\theta,\ \zeta^2=0$. The next three dark KdV systems read
\begin{equation}\label{DKdVs4}
\left\{\begin{array}{l}
u_{t}=J_{1x},\\
v_{i,t}=\hat{Y}v_i,\ i=1,\ 2,\\
v_{12t}=\hat{K}v_{12}+3(v_2v_{1x}-v_1v_{2x})_x,
\end{array}
\right.
\end{equation}
\begin{equation}\label{DKdVs8}
\left\{\begin{array}{l}
u_{t}=J_{1x},\\
v_{it}=\hat{Y}v_i,\ i=1,\ 2,\ 3\\
v_{ijt}=\hat{K}v_{ij}+3(v_jv_{ix}-v_iv_{jx})_x, \ i,j=1,\ 2,\ 3, j>i,\\
v_{123t}=\hat{Y}v_{123}+3(v_1v_{23}-v_2v_{13}+v_3{v_{12}})_x,
\end{array}
\right.
\end{equation}
and
\begin{equation}\label{DKdVs16}
\left\{\begin{array}{l}
u_{t}=J_{1x},\\
v_{it}=\hat{Y}v_i,\ i=1,\ 2,\ 3,\ 4,\\
v_{ijt}=\hat{K}v_{ij}+3(v_{ix}v_j-v_iv_{jx})_x, \ i,\ j=1,\ 2,\ 3,\ 4,\ j>i,\\
v_{ijkt}=\hat{Y}v_{ijk}+3(v_iv_{jk}-v_jv_{ik}+v_{ij}v_k)_x,\ i,\ j,\ k=1,\ 2,\ 3,\ 4,\ i<j<k,\\
v_{1234t}=\hat{K}v_{1234}+6(v_{12}v_{34}+v_{14}v_{23}-v_{13}v_{24})_x+3(v_{123}v_{4x}\\
\qquad\qquad-v_{123x}v_4+v_{124}v_{3x}
-v_{124x}v_3+v_{134}v_{2x}-v_{134x}v_2+v_{234}v_{1x}-v_{234x}v_1)_x,
\end{array}
\right.
\end{equation}
where the operators $\hat{Y}$ and $\hat{K}$ are defined in \eqref{type4} and \eqref{sym}, respectively.

The exact solutions of the supersymmetric KdV equation \eqref{sKdV} are related to the dark KdV systems by $\Phi=\xi+w\theta$ with
\begin{equation}\label{Ds3}
\xi=v_1\zeta_1+v_2\zeta_2,\ w=u+v_{12}\zeta_1\zeta_2
\end{equation}
for \eqref{DKdVs4},
\begin{equation}\label{Ds4}
\xi=v_1\zeta_1+v_2\zeta_2+v_3\zeta_3+v_{123}\zeta_1\zeta_2\zeta_3,\
w=u+v_{12}\zeta_1\zeta_2+v_{13}\zeta_1\zeta_3+v_{23}\zeta_2\zeta_3
\end{equation}
for \eqref{DKdVs8} and
\begin{eqnarray}\label{wxi}
&&w=u+v_{12}\zeta_1\zeta_2+v_{13}\zeta_1\zeta_3+v_{14}\zeta_1\zeta_4+v_{23}\zeta_2\zeta_3+v_{24}\zeta_2\zeta_4
+v_{34}\zeta_3\zeta_4+v_{1234}\zeta_1\zeta_2\zeta_3\zeta_4,\nonumber\\
&&\xi=v_1\zeta_1+v_2\zeta_2+v_3\zeta_3+v_4\zeta_4
+v_{123}\zeta_1\zeta_2\zeta_3+v_{124}\zeta_1\zeta_2\zeta_4
+v_{134}\zeta_1\zeta_3\zeta_4+v_{234}\zeta_2\zeta_3\zeta_4
\end{eqnarray}
for \eqref{DKdVs16}, respectively, where $\zeta_i,\ i=1,\ 2,\ 3,\ 4$ are four different Grassmann parameters and
$u,\ v_i,\ v_{ij},\ v_{ijk}, i<j<k<l,\ i,\ j,\ k=1,\ 2,\ 3,\ 4$ and $v_{1234}$ are sixteen dark boson fields.

Another simple integrable supersymmetric fermionic KdV equation possesses the form
\begin{equation}\label{sBKdV}
\Phi_t=-\Phi_{xxx}+6({\cal D}\Phi)\Phi_x,\ \Phi=\xi+w\theta.
\end{equation}
Utilizing the bosonization method \eqref{wxi} on the supersymmetric KdV equation \eqref{sBKdV}, it is possible to derive a series of integrable dark KdV systems
\begin{eqnarray}\label{sBKdVb}
&&\left\{\begin{array}{l}
u_t=J_{1x},\\
v_{it}=\tilde{K}v_i,\ \tilde{K}\equiv (6u-\partial_x^2)\partial_x, \\
v_{ijt}=\hat{K}v_{ij},\ i,\ j=1,\ 2,\ 3,\ 4,\ i<j,\\
v_{ijkt}=\tilde{K}v_{ijk}+6(v_{ix}v_{jk}-v_{jx}v_{ik}+v_{ij}v_{kx}),\ i<j<k,\ i,\ j,\ k=1,\ 2,\ 3,\ 4,\\
v_{1234t}=\hat{K}v_{1234}+6(v_{12}v_{34}-v_{13}v_{24}+v_{14}v_{23})_x,
\end{array}\right.
\end{eqnarray}
where the operators $\hat{K}$ and $\tilde{K}$ are same as those defined in \eqref{sym} and \eqref{cym}. The simplest trivial case of \eqref{sBKdVb} with $v_2=v_3=v_4=v_{ij}=v_{ijk}=v_{1234}=0$ and $v_1\rightarrow v$ is just the dark KdV system \eqref{type3} for $a=b=0$.

The triviality in these examples comes from the fact that the resulting systems are linear in the fermionic field and, as a result, the first equation of these systems is simply the KdV equation itself.

The integrability of the dark systems \eqref{DKdVs4}, \eqref{DKdVs8}, \eqref{DKdVs16}, and \eqref{sBKdVb} is evident as they emerge as specific reductions of the integrable supersymmetric KdV systems \eqref{sKdV} and \eqref{sBKdV}. Notably, the higher order symmetries of \eqref{DKdVs4}, \eqref{DKdVs8}, \eqref{DKdVs16}, and \eqref{sBKdVb} can be derived by employing the same bosonization procedure applied to the higher order supersymmetry KdV equations \cite{LiuTian} associated with the supersymmetric KdV equations \eqref{sKdV} and \eqref{sBKdV}.

\section{Dark KdV systems from supersymmetric integrable boson systems}

By exploring different variants of supersymmetric KdV equations and coupled KdV systems, one can uncover a broader range of integrable dark systems.

For instance, the supersymmetric bosonic KdV equation,
\begin{equation}\label{BKdV}
U_t=(3U^2-U_{xx})_x,\ U(x,\ t,\ \theta)=w(x,\ t)+\xi(x,\ t)\theta,
\end{equation}
also allows infinitely many dark KdV systems. One of the simple integrable dark KdV system reads
\begin{eqnarray}\label{BKdVb}
&&\left\{\begin{array}{l}
u_t=J_{1x},\\
v_{it}=\hat{K}v_i,\ i=1,\ 2,\ 3,\ 4, \\
v_{ijt}=\hat{K}v_{ij},\ i,\ j=1,\ 2,\ 3,\ 4,\ i<j,\\
v_{ijkt}=\hat{K}v_{ijk}+6(v_iv_{jk}-v_jv_{ik}+v_{ij}v_k)_x,\ i<j<k,\ i,\ j,\ k=1,\ 2,\ 3,\ 4,\\
v_{1234t}=\hat{K}v_{1234}+6(v_{12}v_{34}-v_{13}v_{24}+v_{14}v_{23})_x
\end{array}\right.
\end{eqnarray}
with sixteen dark boson fields $\{u,\ v_{i},\ v_{ij},\ v_{ijk},\ v_{ijkl},\ i, j, k, l=1,\ \ldots,\ 4,\ i<j<k<l\}$.

The dark KdV system \eqref{BKdVb} is related to the supersymmetric KdV equation \eqref{BKdV} by
$U=w+\xi\theta$ with the same $w$ and $\xi$ given in \eqref{wxi}.

The generalized supersymmetric boson KdV system possesses the form (in potential form)
\begin{equation}\label{gBKdV}
U_t=3U_x^2+c({\cal{D}}U_x)({\cal{D}}U)-U_{xxx},\ U(x,\ t,\ \theta)=p(x,\ t)+\xi(x,\ t)\theta.
\end{equation}
In terms of the component fields, \eqref{gBKdV} is equivalent to
\begin{equation}\label{gBKdVc}
\left\{\begin{array}{l}
w_t=(3w^2-w_{xx})_x+c\xi_{xx}\xi,\quad w=p_x, \\
\xi_t=(cw\xi-\xi_{xx})_x+2(3-c)w\xi_x.
\end{array}\right.
\end{equation}
Applying the bosonization formulae \eqref{wxi} to \eqref{gBKdVc}, we have
\begin{eqnarray}\label{BKdVbc}
&&\left\{\begin{array}{l}
u_t=J_{1x},\\
v_{it}=\hat{K}_1v_i,\ \hat{K}_1=cu_x+(6-c)u\partial_x-\partial_x^3,\\
v_{ijt}=\hat{K}v_{ij}+c(v_{ix}v_j-v_iv_{jx})_x,\ i,\ j,\ k=1,\ 2,\ 3,\ 4,\ i<j<k,\\
v_{ijkt}=\hat{K}_1v_{ijk}+(6-c)(v_{ix}v_{jk}-v_{jx}v_{ik}+v_{ij}v_{kx})+c(v_iv_{jkx}-v_jv_{ikx}+v_{ijx}v_k),\\
v_{1234t}=\hat{K}v_{1234}+6(v_{12}v_{34}-v_{13}v_{24}+v_{14}v_{23})_x+c(v_{123x}v_{4}\\
\qquad\qquad -v_{123}v_{4x}-v_{124x}v_{3}+v_{124}v_{3x}+v_{134x}v_{2}-v_{134}v_{2x}
-v_{234x}v_{1}+v_{234}v_{1x})_x.
\end{array}\right.
\end{eqnarray}
The special cases of $c=3$ and $c=0$ hold particular significance in the context under consideration.
When $c=3$, it is noteworthy that $\hat{K}_1=\hat{Y}$, and equation \eqref{BKdVbc} is found to be identical to equation \eqref{DKdVs16}. This observation leads to the implication that the supersymmetric boson field equation \eqref{gBKdV} with $c=3$ is indeed equivalent to the supersymmetric fermion field equation \eqref{sKdV}.
Similarly, for the case of $c=0$, it is highlighted that $\hat{K}_1=\tilde{K}$, resulting in the reduction of equation \eqref{BKdVbc} back to equation \eqref{sBKdVb}. These findings reveal significant relationships and equivalences between the equations for different values of $c$. The implications of these relationships are contingent upon the specific mathematical and physical context in which these equations are being examined.

\section{Dark KdV systems from the bosonization of the superintegrable  KdV system}

The extension of the bosonization technique broadens its applicability and offers a valuable framework for studying diverse fermionic systems, including the super-integrable KdV system \cite{Super,Super1,Super2}

\begin{equation}\label{superKdV}
\left\{\begin{array}{l}
w_t=(3w^2-w_{xx}+3\xi\xi_{x})_x,\\
\xi_t=-4\xi_{xxx}+6w\xi_x+3w_x\xi,
\end{array}\right.
\end{equation}
where $\xi$ is a fermion field and $w$ is a boson field.

Applying the same bosonization solution \eqref{wxi} to the super-integrable KdV system \eqref{superKdV},
we get a new type of integrable dark KdV system,
\begin{eqnarray}\label{BKdVsup}
&&\left\{\begin{array}{l}
u_t=J_{1x},\\
v_{it}=\hat{L}v_i, \ \hat{L}\equiv -4\partial_x^3+6u\partial_x+3u_x,\ i=1,\ 2,\ 3,\ 4,\\
v_{ijt}=\hat{K}v_{ij}+3(v_iv_{jx}-v_{ix}v_j)_x,\ i,\ j,\ k=1,\ 2,\ 3,\ 4,\ i<j<k,\\
v_{ijk}=\hat{L}v_{ijk}+6(v_{jk}v_{ix}-v_{ik}v_{jx}+v_{ij}v_{kx})+3(v_{jkx}v_{x}-v_{ikx}v_{j}+v_{ijx}v_{k})
,\\
v_{1234t}=\hat{K}v_{1234}+6(v_{12}v_{34}-v_{13}v_{24}+v_{14}v_{23})_x+3(v_1v_{234x}-v_{1x}v_{234}\\
\qquad\qquad
-v_2v_{134x}+v_{2x}v_{134}+v_3v_{124x}-v_{3x}v_{124}-v_4v_{123x}+v_{4x}v_{123})_x.
\end{array}\right.
\end{eqnarray}
Taking $v_2=v_3=v_4=0,\ v_{ij}=v_{ijk}=0, i,\ j,\ k=1,\ 2,\ 3,\ 4$ and $v_{1234}=0$, the dark KdV system \eqref{BKdVsup} is reduced back to the two-component dark system \eqref{type1} with $a=0$.

A four-component integrable dark KdV system and an eight-component integrable dark KdV system can be simply read off from \eqref{BKdVsup} by taking $v_3=v_4=v_{13}=v_{14}=v_{23}=v_{24}=v_{34}=v_{123}=v_{124}=v_{134}=v_{234}=v_{1234}=0$ and $v_4=v_{14}=v_{24}=v_{34}=v_{124}=v_{134}=v_{234}=v_{1234}=0$, respectively.

\section{Nonlinear dark KdV systems}
In this section, we extend the linear dark systems \eqref{XE} to nonlinear systems \eqref{XE1} for the KdV equation. Moreover, when studying higher-dimensional integrable systems such as the (2+1)-dimensional KP, BKP, and CKP equations, it is feasible to identify diverse lower-dimensional integrable systems, some of which may exhibit characteristics akin to (extended) dark equations.
For instance, to solve the BKP equation \cite{HaoLou}
\begin{equation}\label{BKP}
v_{xt}+(v_{xxxx}+15vv_{xx}+15v^3-15vq-5v_{xy})_{xx}-5v_{yy},\ q_x=v_y,
\end{equation}
one can find the following nonlinear dark KdV system
\begin{equation}\label{KdV-BKP}
\left\{\begin{array}{l}
u_y=\left(u_{xx}-\frac{a}2u^2\right)_x, \\
v_y=\left(v_{xx}+3v^2+auv+\frac{a^2}6\right)_x
\end{array}\right.
\end{equation}
with the fifth order flow ($w=au+3v$),
\begin{equation}\label{KdV5-BKP}
\left\{\begin{array}{l}
u_t=\left(9u_{xxxx}-15auu_{xx}-\frac{15a}2u_x^2+\frac{5a^2}2u^3\right)_x, \\
v_t=\left[9v_{xxxx}+15(w+3v)v_{xx}+5awu_{xx}+15w_xv_x+\frac{15v}2(w^2+3v^2)
+\frac{5a^2}2u_x^2\right]_x.
\end{array}\right.
\end{equation}
A direct proof can establish that if $u$ and $v$ are solutions of the dark KdV system \eqref{KdV-BKP}, then $v$ satisfies the BKP equation \eqref{BKP}.
In Refs. \cite{HL, HaoLou}, it has been established that when $a=-3$ or $a=-6$, both $u$ and $v$ serve as solutions to the BKP equation \eqref{BKP}.

\section{Supersymmetric dark KdV systems}

The previous sections (sections 5 and 6) have revealed that within a single integrable supersymmetric model, the bosonization method allows for the identification of diverse dark fields. These dark fields are characterized by satisfying graded linear dark equations.  In this section, we delve deeper into the investigation of supersymmetric dark KdV systems.

The most general possible supersymmetric extension of the KdV equation is a one-parameter family of equations with only one fermionic field and with no inverse power of the fields
\begin{equation}\label{GSKdV}
\Phi_t=-\Phi_{xxx}+a({\cal{D}}\Phi_x)\Phi+(6-a)({\cal{D}}\Phi)\Phi_x.
\end{equation}
A comprehensive Painlev\'e analysis indicates that within the class of one-parameter systems, only two distinctive cases, specifically equation \eqref{sKdV} for $a=3$ and equation \eqref{sBKdV} for $a=0$, exhibit integrability. Notably, an additional significant characteristic of these two equations is their ability to be formulated as Hamiltonian systems, employing the Poisson brackets associated with the second Hamiltonian structure \cite{SuperKdV}.

\subsection{Supersymmetric dark KdV systems related to the supersymmetric KdV equation (50)}

	To construct the most comprehensive supersymmetric dark KdV system, in a manner akin to the classical integrable cases, it is necessary to consider a linear combination of all possible terms which ensures that all relevant terms and interactions are taken into account
\begin{equation}\label{DGSKdV}
\left\{\begin{array}{l}
\Phi_t=-\Phi_{xxx}+3[({\cal{D}}\Phi)\Phi]_x,\\
\Psi_t=-a_1\Psi_{xxx}+a_2\Phi{\cal{D}}\Psi_x+a_3({\cal{D}}\Phi_x)\Psi+a_4\Phi_x{\cal{D}}\Psi
+a_5({\cal{D}}\Phi)\Psi_x
\end{array}\right.
\end{equation}
such that it is symmetry integrable.

Upon completing arduous computations, one can ascertain the existence of supersymmetric integrable dark KdV systems characterized by the presence of higher order symmetries.

\bf Case 1. \rm The first type of supersymmetric dark KdV systems is characterized by a specific form
\begin{equation}\label{DGSKdV1}
\left\{\begin{array}{l}
\Phi_t=-\Phi_{xxx}+3(\Phi{\cal{D}}\Phi)_x,\\
\Psi_t=-\Psi_{xxx}+3\Phi{\cal{D}}\Psi_x,
\end{array}\right.
\end{equation}
where $\Psi$ is also a fermionic super field. The symmetry integrability of \eqref{DGSKdV1} is guaranteed by the existence of higher order symmetry
\begin{equation}\label{5SKdV1}
\left\{\begin{array}{l}
\Phi_{\tau}=-\Phi_{xxxxx}+\big[5(\Phi_{x}{\cal{D}}\Phi)_x+5\Phi{\cal{D}}\Phi_{xx}
-10\Phi({\cal{D}}\Phi)^2\big]_x,\\
\Psi_{\tau}=-\Psi_{xxxxx}+5(\Phi {\cal{D}}\Psi_{xx})_x +5(\Phi_{xx}-2\Phi{\cal{D}}\Phi) {\cal{D}}\Psi_{x}.
\end{array}\right.
\end{equation}
In the supersymmetric case, there are two types of conservation laws. One possesses the same forms
\begin{equation}\label{CL1}
\rho_t=J_x,
\end{equation}
as those for the usual integrable systems, and the other can be written as
\begin{equation}\label{CL2}
\rho_t={\cal{DF}},
\end{equation}
where $\rho$ is the conserved density while $J$ and ${\cal{F}}$ are the conserved flows related to the first and second types of conservation laws, respectively.
It is clear that $\rho=\Phi$ and $J=-\Phi_{xx}+3\Phi{\cal{D}}\Phi$ is related to the first type of the conservation laws \eqref{CL1}. From \eqref{DGSKdV1}, one can also check that
\begin{equation}\label{CL21}
\rho=\Psi_x\Phi,\ {\cal{F}}={\cal{D}}(\Psi_{xx}\Phi_x-\Psi_x\Phi_{xx}-\Psi_{xxx}\Phi)+3\Phi\Phi_x\Psi_x
\end{equation}
are related to the second type of conservation laws \eqref{CL2}.\\
\bf Case 2. \rm
The second supersymmetric dark KdV equation reads
\begin{equation}\label{DGSKdV1a}
\left\{\begin{array}{l}
\Phi_t=-\Phi_{xxx}+3(\Phi{\cal{D}}\Phi)_x,\\
\Psi_t=-\Psi_{xxx}+3 \Psi_x{\cal{D}}\Phi,
\end{array}\right.
\end{equation}
with the fifth order supersymmetric flow
\begin{equation}\label{5SKdV1a}
\left\{\begin{array}{l}
\Phi_{\tau}=-\Phi_{xxxxx}+\big[5(\Phi_{x}{\cal{D}}\Phi)_x+5\Phi{\cal{D}}\Phi_{xx}
-10\Phi({\cal{D}}\Phi)^2\big]_x,\\
\Psi_{\tau}=-\Psi_{xxxxx}+5 (\Psi_{xx}{\cal{D}}\Phi)_x +5[{\cal{D}}\Phi_{xx}-2({\cal{D}}\Phi)^2+\Phi\Phi_x] \Psi_{x}.
\end{array}\right.
\end{equation}
\bf Case 3. \rm The third supersymmetric dark KdV system can be expressed as
\begin{equation}\label{DGSKdV2}
\left\{\begin{array}{l}
\Phi_t=-\Phi_{xxx}+3(\Phi{\cal{D}}\Phi)_x,\\
\Psi_t=-\Psi_{xxx}+3\Phi{\cal{D}}\Psi_x+3 \Psi_x{\cal{D}}\Phi.
\end{array}\right.
\end{equation}
In this case, one can prove that $\rho=\Psi\Phi$ is a conserved density with the first type of conservation flow \eqref{CL1},
 $$(\Psi\Phi)_t=(-\Psi_{xx}\Phi-\Psi\Phi_{xx}+\Psi_x\Phi_x+\Psi\Phi{\cal{D}}\Phi)_x.$$

 The integrability of \eqref{DGSKdV2} is guaranteed by the existence of higher order symmetries, for instance, the fifth order symmetry possesses the form
\begin{equation}\label{5SKdV2}
\left\{\begin{array}{l}
\Phi_{\tau}=-\Phi_{xxxxx}+5\big[(\Phi_{x}{\cal{D}}\Phi)_x+\Phi{\cal{D}}\Phi_{xx}
-2\Phi({\cal{D}}\Phi)^2\big]_x,\\
\Psi_{\tau}=-\Psi_{xxxxx}+5(\Phi {\cal{D}}\Psi_{xx}+\Psi_{xx} {\cal{D}}\Phi)_x +5[{\cal{D}}\Phi_{xx}-2({\cal{D}}\Phi)^2] \Psi_{x} +5(\Phi_{xx}-4\Phi{\cal{D}}\Phi) {\cal{D}}\Psi_{x}.
\end{array}\right.
\end{equation}
\bf Case 4. \rm The fourth type of supersymmetric dark KdV systems has the following simple form
\begin{equation}\label{DGSKdV2a}
\left\{\begin{array}{l}
\Phi_t=-\Phi_{xxx}+3(\Phi{\cal{D}}\Phi)_x,\\
\Psi_t=-\Psi_{xxx}+3{\cal{D}}(\Phi \Psi_x)
\end{array}\right.
\end{equation}
accompanied by the fifth order supersymmetric flow
\begin{equation}\label{5SKdV2a}
\left\{\begin{array}{l}
\Phi_{\tau}=-\Phi_{xxxxx}+5\big[(\Phi_{x}{\cal{D}}\Phi)_x+\Phi{\cal{D}}\Phi_{xx}
-2\Phi({\cal{D}}\Phi)^2\big]_x,\\
\Psi_{\tau}=-\Psi_{xxxxx}+5 {\cal{D}}[\Phi\Psi_{xxx}+(\Phi_x\Psi_{x})_x-2\Phi({\cal{D}}\Phi) \Psi_{x}].
\end{array}\right.
\end{equation}
In this case the dark field $\Psi$ itself is a conserved density with the second type of flow ${\cal{F}}=3\Phi \Psi_x-{\cal{D}}\Psi_{xx}$.\\
\bf Case 5. \rm
The fifth supersymmetric dark KdV system
\begin{equation}\label{DGSKdV3}
\left\{\begin{array}{l}
\Phi_t=-\Phi_{xxx}+3(\Phi{\cal{D}}\Phi)_x,\\
\Psi_t=-\Psi_{xxx}+3(\Phi{\cal{D}}\Psi)_x
\end{array}\right.
\end{equation}
possesses the fifth order symmetry,
\begin{equation}\label{5SKdV3}
\left\{\begin{array}{l}
\Phi_{\tau}=-\Phi_{xxxxx}+5\big[(\Phi_{x}{\cal{D}}\Phi)_x+\Phi{\cal{D}}\Phi_{xx}
-2\Phi({\cal{D}}\Phi)^2\big]_x,\\
\Psi_{\tau}=-\Psi_{xxxxx}+5[(\Phi{\cal{D}}\Psi_{x})_x+\Phi_{xx}{\cal{D}}\Psi
-2\Phi({\cal{D}}\Psi)({\cal{D}}\Phi)]_x.
\end{array}\right.
\end{equation}
In the previous four cases, \eqref{DGSKdV1}, \eqref{DGSKdV1a}, \eqref{DGSKdV2}, and \eqref{DGSKdV2a}, the dark field $\Psi$ is not a first type of conserved density. \eqref{DGSKdV3} is the first case with the first type of conserved flow $J=3\Phi{\cal{D}}\Psi-\Psi_{xx}$.\\
\bf Case 6. \rm The sixth type supersymmetric dark KdV system
\begin{equation}\label{C13a}
\left\{\begin{array}{l}
\Phi_t=-\Phi_{xxx}+3(\Phi{\cal{D}}\Phi)_x,\\
\Psi_t=-\Psi_{xxx}+3\big(\Psi{\cal{D}}\Phi\big)_x
\end{array}\right.
\end{equation}
represents a conserved form for the dark field $\Psi$ and its associated fifth order symmetry flow reads
\begin{equation}\label{C13b}
\left\{\begin{array}{l}
\Phi_{\tau}=-\Phi_{xxxxx}+\big[5\Phi_{xx}{\cal{D}}\Phi+5\Phi_x{\cal{D}}\Phi_x+5\Phi{\cal{D}}\Phi_{xx}
-10\Phi({\cal{D}}\Phi)^2\big]_x,\\
\Psi_{\tau}=5\big[\Psi{\cal{D}}\Phi_{xx}+\Psi_x{\cal{D}}\Phi_x
+\Psi_{xx}{\cal{D}}\Phi+\Phi\Phi_x\Psi-2({\cal{D}}\Phi)^2\Psi\big]_x-\Psi_{xxxxx}.
\end{array}\right.
\end{equation}
In this case, the dark field $\Psi$ is also a first type of conserved density with the conserved flow $J=-\Psi_{xx}+3\Psi{\cal{D}}\Phi$.\\
\bf Case 7. \rm The seventh supersymmetric dark KdV system can be written as
\begin{equation}\label{DGSKdV3a}
\left\{\begin{array}{l}
\Phi_t=-\Phi_{xxx}+3(\Phi{\cal{D}}\Phi)_x,\\
\Psi_t=-\Psi_{xxx}+3 {\cal{D}}[({\cal{D}}\Phi)({\cal{D}}\Psi)],
\end{array}\right.
\end{equation}
along with the fifth order supersymmetric flow
\begin{equation}\label{5SKdV3a}
\left\{\begin{array}{l}
\Phi_{\tau}=-\Phi_{xxxxx}+5\big[(\Phi_{x}{\cal{D}}\Phi)_x+\Phi{\cal{D}}\Phi_{xx}
-2\Phi({\cal{D}}\Phi)^2\big]_x,\\
\Psi_{\tau}=-\Psi_{xxxxx}+5 {\cal{D}}\big\{[({\cal{D}}\Phi){\cal{D}}\Psi_{x}]_x
+[{\cal{D}}\Phi_{xx}+\Phi\Phi_x-2({\cal{D}}\Phi)^2]{\cal{D}}\Psi\big\}.
\end{array}\right.
\end{equation}
In this case, the dark field $\Psi$ is a conserved density with the second type of conserved flow ${\cal{F}}=-{\cal{D}}\Psi_{xx}+3 ({\cal{D}}\Phi){\cal{D}}\Psi$.\\
\bf Case 8. \rm The eighth type of supersymmetric KdV systems
\begin{equation}\label{DGSKdV3b}
\left\{\begin{array}{l}
\Phi_t=-\Phi_{xxx}+3(\Phi{\cal{D}}\Phi)_x,\\
\Psi_t=-\Psi_{xxx}+3 \Psi_x{\cal{D}}\Phi+3{\cal{D}}(\Phi\Psi_x).
\end{array}\right.
\end{equation}
possesses a second type of conservation laws
$$(\Psi\Phi_x)_t={\cal{D}}\left[{\cal{D}}\left(\Phi_{xxx}\Psi-\Psi_{xx}\Phi_x
+\Psi_x\Phi_{xx}+3\Psi\Phi{\cal{D}}\Phi_x+3\Psi\Phi_x{\cal{D}}\Phi\right)-3\Phi\Phi_x\Psi_x\right]. $$
The fifth order symmetry flow of \eqref{5SKdV3a} reads
\begin{equation}\label{5SKdV3b}
\left\{\begin{array}{l}
\Phi_{\tau}=5\big[(\Phi_{xx}{\cal{D}}\Phi)_x+\Phi{\cal{D}}\Phi_{xx}
-2\Phi({\cal{D}}\Phi)^2\big]_x-\Phi_{xxxxx},\\
\Psi_{\tau}=5 (2\Psi_{xx} {\cal{D}}\Phi-\Phi{\cal{D}}\Psi_{xx})_x-5(\Phi_{xx}-4\Phi{\cal{D}}\Phi) {\cal{D}}\Psi_{x}\\
\qquad\quad +10[{\cal{D}}\Phi_{xx}-3({\cal{D}}\Phi)^2+2\Phi\Phi_x] \Psi_{x}-\Psi_{xxxxx}.
\end{array}\right.
\end{equation}
\bf Case 9. \rm For the ninth type of supersymmetric dark KdV equation systems
\begin{equation}\label{DGSKdV4}
\left\{\begin{array}{l}
\Phi_t=-\Phi_{xxx}+3(\Phi{\cal{D}}\Phi)_x,\\
\Psi_t=-\Psi_{xxx}+3(\Phi{\cal{D}}\Psi)_x+3{\cal{D}}[({\cal{D}}\Phi)({\cal{D}}\Psi)],
\end{array}\right.
\end{equation}
the dark field $\Psi$ is conserved with the second type of conserved flow
${\cal{F}}= {\cal{D}}(3\Phi{\cal{D}}\Psi-\Psi_{xx})+3({\cal{D}}\Phi){\cal{D}}\Psi$. The fifth order symmetry flow of \eqref{DGSKdV4} possesses the form
\begin{equation}\label{5SKdV4}
\left\{\begin{array}{l}
\Phi_{\tau}=-\Phi_{xxxxx}+5\big[(\Phi_{x}{\cal{D}}\Phi)_x+\Phi{\cal{D}}\Phi_{xx}
-2\Phi({\cal{D}}\Phi)^2\big]_x,\\
\Psi_{\tau}=-\Psi_{xxxxx}+5[\Phi{\cal{D}}\Psi_{xx}+\Psi_{xx}{\cal{D}}\Phi
+\Phi_x{\cal{D}}\Psi_{x}+2\Phi_{xx}{\cal{D}}\Psi-4({\cal{D}}\Psi)({\cal{D}}\Phi)\Phi]_x\\
\qquad \quad
+5{\cal{D}}[\Phi_{xx}\Psi_x-\Phi_x\Psi_{xx}
+({\cal{D}}\Phi_x)({\cal{D}}\Psi_x)-2({\cal{D}}\Phi)^2{\cal{D}}\Psi].
\end{array}\right.
\end{equation}
\bf Case 10. \rm The tenth type of supersymmetric dark KdV systems
\begin{equation}\label{DGSKdV4a}
\left\{\begin{array}{l}
\Phi_t=-\Phi_{xxx}+3(\Phi{\cal{D}}\Phi)_x,\\
\Psi_t=-\Psi_{xxx}+3{\cal{D}}[\Phi\Psi_x+({\cal{D}}\Phi)({\cal{D}}\Psi)]
\end{array}\right.
\end{equation}
is symmetry integrable because of the existence of the higher order symmetries, say,
\begin{equation}\label{5SKdV4a}
\left\{\begin{array}{l}
\Phi_{\tau}=-\Phi_{xxxxx}+5\big[(\Phi_{x}{\cal{D}}\Phi)_x+\Phi{\cal{D}}\Phi_{xx}
-2\Phi({\cal{D}}\Phi)^2\big]_x,\\
\Psi_{\tau}=-\Psi_{xxxxx}+5 (\Psi_{xx} {\cal{D}}\Phi+\Psi_{x} {\cal{D}}\Phi_x)_x-5{\cal{D}}[\Phi\Psi_{xxx}\\
\qquad\quad-({\cal{D}}\Phi_{xx}){\cal{D}}\Psi
+4\Phi({\cal{D}}\Phi)\Psi_x+2({\cal{D}}\Phi)^2{\cal{D}}\Psi].
\end{array}\right.
\end{equation}
In this case, the dark field $\Psi$ is also conserved with the second type of conserved flow
$${\cal{F}}=-{\cal{D}}\Psi_{xx}+3\Phi\Psi_x+3({\cal{D}}\Phi){\cal{D}}\Psi.$$
\bf Case 11. \rm The validity of the eleventh type of supersymmetric dark systems
\begin{equation}\label{DGSKdV5}
\left\{\begin{array}{l}
\Phi_t=-\Phi_{xxx}+3(\Phi{\cal{D}}\Phi)_x,\\
\Psi_t=-\Psi_{xxx}+3(\Phi{\cal{D}}\Psi+\Psi{\cal{D}}\Phi)_x
\end{array}\right.
\end{equation}
is naturally established due to the direct correspondence between the dark field $\Psi$ equation  (the second equation of \eqref{DGSKdV5}) and the symmetry equation of the $\Phi$ equation (the first equation of \eqref{DGSKdV5}). This correspondence indicates that the dark field $\Psi$ serves as a conserved density, accompanied by the first type of conserved flow $J=-\Psi_{xx}+3\Phi{\cal{D}}\Psi+3\Psi{\cal{D}}\Phi$.

The higher order equations for the dark field $\Psi$ are also the symmetry equations of the higher order supersymmetric KdV equations, say, the fifth order symmetry flow of \eqref{DGSKdV5} reads
\begin{equation}\label{5SKdV5}
\left\{\begin{array}{l}
\Phi_{\tau}=-\Phi_{xxxxx}+5\big[(\Phi_{x}{\cal{D}}\Phi)_x+\Phi{\cal{D}}\Phi_{xx}
-2\Phi({\cal{D}}\Phi)^2\big]_x,\\
\Psi_{\tau}=5[(\Psi{\cal{D}}\Phi_{x}+\Phi{\cal{D}}\Psi_{x})_x+ \Psi_{xx}{\cal{D}}\Phi+\Phi_{xx}{\cal{D}}\Psi
-4\Phi({\cal{D}}\Phi){\cal{D}}\Psi-2({\cal{D}}\Phi)^2\Psi]_x-\Psi_{xxxxx}.
\end{array}\right.
\end{equation}
\bf Case 12. \rm The twelfth type of supersymmetric dark KdV systems
\begin{equation}\label{C11a}
\left\{\begin{array}{l}
\Phi_t=-\Phi_{xxx}+3(\Phi{\cal{D}}\Phi)_x,\\
\Psi_t=-\Psi_{xxx}+3(\Psi{\cal{D}}\Phi-\Phi{\cal{D}}\Psi)_x,
\end{array}\right.
\end{equation}
closely resembles the eleventh type \eqref{DGSKdV5}, except for a change in the sign of the last term. The dark field $\Psi$ is also a conserved density with the first type of conserved flow
$J=-\Psi_{xx}+3\Psi{\cal{D}}\Phi-3\Phi{\cal{D}}\Psi$.

The fifth order symmetry flow of \eqref{C11a} can be written as
\begin{equation}\label{C11b}
\left\{\begin{array}{l}
\Phi_{\tau}=-\Phi_{xxxxx}+5\big[(\Phi_{x}{\cal{D}}\Phi)_x+\Phi{\cal{D}}\Phi_{xx}
-2\Phi({\cal{D}}\Phi)^2\big]_x,\\
\Psi_{\tau}=5\big[(\Psi{\cal{D}}\Phi_{x}-\Phi_{x}{\cal{D}}\Psi)_x
+\Psi_{xx}{\cal{D}}\Phi-\Phi{\cal{D}}\Psi_{xx}+2\Phi\Phi_x\Psi\\
\qquad \quad
+2\Phi({\cal{D}}\Phi){\cal{D}}\Psi-2({\cal{D}}\Phi)^2\Psi\big]_x-\Psi_{xxxxx}.
\end{array}\right.
\end{equation}
\bf Case 13. \rm The thirteenth supersymmetric dark KdV system,
\begin{equation}\label{C12a}
\left\{\begin{array}{l}
\Phi_t=-\Phi_{xxx}+3(\Phi{\cal{D}}\Phi)_x,\\
\Psi_t=-\Psi_{xxx}+3\big(-\Phi{\cal{D}}\Psi+2\Psi{\cal{D}}\Phi\big)_x,
\end{array}\right.
\end{equation}
has a nearly identical form to the twelfth type \eqref{C11a}, except for a numerical factor of ``2" in the last term of \eqref{C12a}. It is clear that the dark field $\Psi$ is conserved with the first type of conserved flow
$J=-\Psi_{xx}-3\Phi{\cal{D}}\Psi+6\Psi{\cal{D}}\Phi$.

The fifth symmetry flow of \eqref{C12a} possesses the form
\begin{equation}\label{C12b}
\left\{\begin{array}{l}
\Phi_{\tau}=-\Phi_{xxxxx}+5\big[(\Phi_{x}{\cal{D}}\Phi)_x+\Phi{\cal{D}}\Phi_{xx}
-2\Phi({\cal{D}}\Phi)^2\big]_x,\\
\Psi_{\tau}=5\big\{(2\Psi{\cal{D}}\Phi_{x}-\Phi_{x}{\cal{D}}\Psi)_x
+2\Psi_{xx}{\cal{D}}\Phi-\Phi{\cal{D}}\Psi_{xx}+4\Phi[\Phi_x\Psi
+({\cal{D}}\Phi){\cal{D}}\Psi]\\
\quad\qquad -6({\cal{D}}\Phi)^2\Psi\big\}_x-\Psi_{xxxxx}.
\end{array}\right.
\end{equation}
\bf Case 14. \rm The fourteenth supersymmetric dark KdV system
\begin{equation}\label{C14a}
\left\{\begin{array}{l}
\Phi_t=-\Phi_{xxx}+3(\Phi{\cal{D}}\Phi)_x,\\
\Psi_t=-4\Psi_{xxx}+{\cal{D}}\big(6\Phi\Psi_x+3\Phi_x\Psi\big),
\end{array}\right.
\end{equation}
possesses the first type of conserved density $\Phi$ and the second type of conserved density $\Psi$ with the conserved flows $J=-\Phi_{xx}+3\Phi{\cal{D}}\Phi$ and ${\cal{F}}=-4\Psi_{xx}+6\Phi\Psi_x+3\Phi_x\Psi$, respectively.

The higher order symmetry flows derived from equation \eqref{C14a} can be expressed as combinations of two distinct types of conservation laws. The first type corresponds to conservation law formulated in terms of the field $\Phi$, while the second type pertains to conservation law formulated in terms of the field $\Psi$, say,
\begin{equation}\label{C14b}
\left\{\begin{array}{l}
\Phi_{\tau}=-\Phi_{xxxxx}+5\big[(\Phi_{x}{\cal{D}}\Phi)_x+\Phi{\cal{D}}\Phi_{xx}
-2\Phi({\cal{D}}\Phi)^2\big]_x,\\
\Psi_{\tau}=5\big[{\cal{D}}(3\Phi_{xx}\Psi+8\Phi\Psi_{xx}+4\Phi_x\Psi_{x})+4\Phi\Phi_x\Psi\big]_x-16\Psi_{xxxxx}
\\
\qquad \quad +5{\cal{D}}\big[3\Phi_{xx}\Psi_x-6(\Phi\Psi)_x{\cal{D}}\Phi-2\Phi\Phi_x{\cal{D}}\Psi\big].
\end{array}\right.
\end{equation}
\bf Case 15. \rm The fifteenth supersymmetric dark KdV system corresponds to
\begin{equation}\label{C15a}
\left\{\begin{array}{l}
\Phi_t=-\Phi_{xxx}+3(\Phi{\cal{D}}\Phi)_x,\\
\Psi_t=-4\Psi_{xxx}+6 \Phi{\cal{D}}\Psi_x+3\Phi_x{\cal{D}}\Psi.
\end{array}\right.
\end{equation}
Regarding the fifth order symmetry flow of equation \eqref{C15a}, it can be expressed as
\begin{equation}\label{C15b}
\left\{\begin{array}{l}
\Phi_{\tau}=-\Phi_{xxxxx}+5\big[(\Phi_{x}{\cal{D}}\Phi)_x+\Phi{\cal{D}}\Phi_{xx}
-2\Phi({\cal{D}}\Phi)^2\big]_x,\\
\Psi_{\tau}=10\Phi\Phi_x \Psi_x+40\Phi{\cal{D}}\Psi_{xxx}+60\Phi_x{\cal{D}}\Psi_{xx}+10(5\Phi_{xx}-3\Phi{\cal{D}}\Phi){\cal{D}}\Psi_x\\ \qquad \quad+5\big(3\Phi_{xxx}-2\Phi_x{\cal{D}}\Phi-4\Phi{\cal{D}}\Phi_x\big){\cal{D}}\Psi-16\Psi_{xxxxx}.
\end{array}\right.
\end{equation}
\bf Case 16. \rm The sixteenth supersymmetric dark KdV system
\begin{equation}\label{DGSKdV6}
\left\{\begin{array}{l}
\Phi_t=-\Phi_{xxx}+3(\Phi{\cal{D}}\Phi)_x,\\
\Psi_t=\Phi_x{\cal{D}}\Psi+2 \Psi_x{\cal{D}}\Phi+b ({\cal{D}}\Phi_x)\Psi,
\end{array}\right.
\end{equation}
with an arbitrary constant $b$, can be described as a first order linear equation for the dark field $\Psi$. Additionally, the higher order symmetry flows derived from equation \eqref{DGSKdV6} can also be expressed as first order linear equations for the dark field $\Psi$, say
\begin{equation}\label{5SKdV6a}
\left\{\begin{array}{l}
\Phi_{\tau}=-\Phi_{xxxxx}+5\big[(\Phi_{x}{\cal{D}}\Phi)_x+\Phi{\cal{D}}\Phi_{xx}
-2\Phi({\cal{D}}\Phi)^2\big]_x,\\
\Psi_{\tau}=2\big[{\cal{D}}\Phi_{xx} -3({\cal{D}}\Phi)^2+2\Phi\Phi_x\big] \Psi_x +\big(\Phi_{xxx}-2\Phi{\cal{D}}\Phi_x-4\Phi_x{\cal{D}}\Phi\big){\cal{D}}\Psi\\
\qquad \quad +b[{\cal{D}}\Phi_{xxx}
+2\Phi\Phi_{xx}-6({\cal{D}}\Phi)({\cal{D}}\Phi_x)]\Psi.
\end{array}\right.
\end{equation}
\bf Case 17. \rm The final case corresponds to a degenerate scenario where the system becomes trivial and completely decoupled
\begin{equation}\label{DGSKdV7}
\left\{\begin{array}{l}
\Phi_t=-\Phi_{xxx}+3(\Phi{\cal{D}}\Phi)_x,\\
\Psi_t=-a \Psi_{xxx}
\end{array}\right.
\end{equation}
In this case, the higher order symmetry flows are also trivial and do not contribute to the system's behavior
\begin{equation}\label{DGSKdV7a}
\left\{\begin{array}{l}
\Phi_t=-\Phi_{xxx}+3(\Phi{\cal{D}}\Phi)_x,\\
\Psi_{t_{2n+1}}=-a_{2n+1}\partial_x^{2n+1} \Psi,\ n=1,\ 2,\ \ldots,\ \infty,
\end{array}\right.
\end{equation}
where $a_{2n+1},\ n=1,\ 2,\ \ldots,\ \infty,$ are arbitrary constants.

\subsection{Supersymmetric integrable dark KdV systems related to the supersymmetric KdV equation (57)}

In this section, we list a comprehensive classification for the supersymmetric dark KdV system in the form of
\begin{equation}\label{SDKdV2}
\left\{\begin{array}{l}
\Phi_t=-\Phi_{xxx}+6\Phi_x{\cal{D}}\Phi,\\
\Psi_t=a_1\Psi_{xxx}+a_2\Phi {\cal{D}}\Psi_x+a_3({\cal{D}}\Phi_x)\Psi+a_4\Phi_x{\cal{D}}\Psi+a_5({\cal{D}}\Phi)\Psi_x
\end{array}\right.
\end{equation}
with the fifth order symmetry flow
\begin{equation}\label{SDKdV2h}
\left\{\begin{array}{l}
\Phi_{\tau}=-\Phi_{xxxxx}+10(\Phi_{xx}{\cal{D}}\Phi)_x+10\Phi_{x}{\cal{D}}\Phi_{xx}
-30\Phi_x({\cal{D}}\Phi)^2,\\
\Psi_{\tau}=b_1\Psi_{xxxxx}+b_2\Phi{\cal{D}}\Psi_{xxx}+b_3({\cal{D}}\Phi)\Psi_{xxx}+b_4\Phi_x{\cal{D}}\Psi_{xx}
+b_5({\cal{D}}\Phi_x)\Psi_{xx}+(b_6\Phi{\cal{D}}\Phi\\
\qquad\quad +b_7\Phi_{xx}){\cal{D}}\Psi_x+[b_8({\cal{D}}\Phi)^2
+b_9{\cal{D}}\Phi_{xx}+b_{10}\Phi\Phi_x]\Psi_x+(b_{11}\Phi_x{\cal{D}}\Phi+b_{12}\Phi{\cal{D}}\Phi_x
\\
\qquad \quad +b_{13}\Phi_{xxx}){\cal{D}}\Psi+[b_{14}({\cal{D}}\Phi){\cal{D}}\Phi_x+b_{15}\Phi\Phi_{xxx}
+b_{16}{\cal{D}}\Phi_{xxx}]\Psi,
\end{array}\right.
\end{equation}
where $a_1,\ a_2,\ \ldots,\ a_5$ and $b_1,\ b_2,\ \ldots,\ b_{16}$ are any possible constants.

	Through the imposition of consistent conditions, specifically $\Phi_{t\tau}=\Phi_{\tau t}$ and $\Psi_{t\tau}=\Psi_{\tau t}$, we are able to ascertain the existence of twelve unique supersymmetric dark KdV systems, each characterized by a distinct equation structure.
\vskip.05in
\leftline{(I.a) $\qquad \quad
\left\{\begin{array}{l}
\Phi_{t}=-\Phi_{xxx}+6\Phi_x{\cal{D}}\Phi,\\
\Psi_t=-4\Psi_{xxx}+6({\cal{D}}\Phi)\Psi_x+3({\cal{D}}\Phi_x)\Psi,
\end{array}\right.
$}\vskip.05in
\leftline{(I.b) $ \qquad \quad
\left\{\begin{array}{l}
\Phi_{\tau}=10(\Phi_{xx}u)_x+10\Phi_{x}u_{xx}
-30\Phi_xu^2-\Phi_{xxxxx}, u\equiv {\cal{D}}\Phi,\\
\Psi_{\tau}=5(8\Psi_{xx}u+3\Psi u_{xx}+4\Psi_x u_x
-6u^2\Psi)_x+15\Psi_x u_{xx}+30uu_x\Psi-16\Psi_{xxxxx},
\end{array}\right.
$}
\noindent which is a supersymmetric form of the usual first type of dark KdV \eqref{type1}. Applying the supersymmetric derivative ${\cal{D}}$ on the first equation of (I.a) and making the transformations,
\begin{equation}
{\cal{D}}\Phi=u,\ \Psi=\zeta v, \label{tr1}
\end{equation}
where $\zeta$ is an arbitrary Grassmann parameter $\zeta$,
one can immediately obtain the usual KdV equation $u_t=-u_{xxx}+6uu_x$ because of $\Phi_x^2=0$ for the fermion field $\Phi$. While the second equation of (I.a) is just the $v$ equation of \eqref{type1}.\vskip.05in

\leftline{(II.a) $ \qquad\qquad
\left\{\begin{array}{l}
\Phi_{t}=-\Phi_{xxx}+6\Phi_x{\cal{D}}\Phi,\\
\Psi_t=(-\Psi_{xx}+6\Psi{\cal{D}}\Phi)_x,
\end{array}\right.
$}\vskip.05in

\leftline{(II.b) $ \qquad\qquad
\left\{\begin{array}{l}
\Phi_{\tau}=-\Phi_{xxxxx}+10(\Phi_{xx}{\cal{D}}\Phi)_x+10\Phi_{x}{\cal{D}}\Phi_{xx}
-30\Phi_x({\cal{D}}\Phi)^2,\\
\Psi_{\tau}=[-\Psi_{xxxx}+10(\Psi_{x}{\cal{D}}\Phi)_x+10\Psi{\cal{D}}\Phi_{xx}-30({\cal{D}}\Phi)^2\Psi]_x.
\end{array}\right.
$}\vskip.05in
This simple case can be considered as the first kind of supersymmetric form of the usual symmetry related dark KdV system \eqref{type2} because the same transformation \eqref{tr1} makes (II.a) becomes \eqref{type2}. \vskip.05in

\leftline{(III.a) $ \qquad\qquad
\left\{\begin{array}{l}
\Phi_{t}=-\Phi_{xxx}+6\Phi_x{\cal{D}}\Phi,\\
\Psi_t=-\Psi_{xxx}+6{\cal{D}}[({\cal{D}}\Phi){\cal{D}}\Psi],
\end{array}\right.
$}\vskip.05in

\leftline{(III.b) $ \qquad\qquad
\left\{\begin{array}{l}
\Phi_{\tau}=-\Phi_{xxxxx}+10(\Phi_{xx}{\cal{D}}\Phi)_x+10\Phi_{x}{\cal{D}}\Phi_{xx}
-30\Phi_x({\cal{D}}\Phi)^2,\\
\Psi_{\tau}=[-\Psi_{xxx}+10(\Phi_x{\cal{D}}\Psi)_x+10\Psi_{xx}{\cal{D}}\Phi]_x
+10{\cal{D}}[\Phi_{xx}\Psi_x-3({\cal{D}}\Phi)^2{\cal{D}}\Psi].
\end{array}\right.
$}\vskip.05in
This case can be considered as the second type of supersymmetric extension of the usual symmetry related dark KdV system because Eq. (III.a) is related to \eqref{type2} by the following transformation
\begin{equation}
{\cal{D}}\Phi=u,\ {\cal{D}}\Psi=v. \label{tr2}
\end{equation}

\leftline{(IV.a) $ \qquad\qquad
\left\{\begin{array}{l}
\Phi_{t}=-\Phi_{xxx}+6\Phi_x{\cal{D}}\Phi,\\
\Psi_t=-\Psi_{xxx}+6({\cal{D}}\Phi)\Psi_x,
\end{array}\right.
$}\vskip.05in

\leftline{(IV.b) $ \qquad\qquad
\left\{\begin{array}{l}
\Phi_{\tau}=-\Phi_{xxxxx}+10(\Phi_{xx}{\cal{D}}\Phi)_x+10\Phi_{x}{\cal{D}}\Phi_{xx}
-30\Phi_x({\cal{D}}\Phi)^2,\\
\Psi_{\tau}=-\Psi_{xxxxx}+10(\Psi_{xx}{\cal{D}}\Phi)_{x}+10[{\cal{D}}\Phi_{xx}-3({\cal{D}}\Phi)^2]\Psi_x.
\end{array}\right.
$}\vskip.05in
By using the transformation \eqref{tr1}, the supersymmetric dark KdV system (IV.a) can be converted to the usual dark KdV system \eqref{type3} related to the dual symmetry of the usual KdV equation. \vskip.05in

\leftline{(V.a) $ \qquad\qquad
\left\{\begin{array}{l}
\Phi_{t}=-\Phi_{xxx}+6\Phi_x{\cal{D}}\Phi,\\
\Psi_t=(-\Psi_{xx}+3\Psi{\cal{D}}\Phi)_x,
\end{array}\right.
$}\vskip.05in
\leftline{(V.b) $ \qquad\qquad
\left\{\begin{array}{l}
\Phi_{\tau}=-\Phi_{xxxxx}+10(\Phi_{xx}{\cal{D}}\Phi)_x+10\Phi_{x}{\cal{D}}\Phi_{xx}
-30\Phi_x({\cal{D}}\Phi)^2,\\
\Psi_{\tau}=[-\Psi_{xxxx}+5(\Psi_{x}{\cal{D}}\Phi)_x+5\Psi{\cal{D}}\Phi_{xx}-10\Psi({\cal{D}}\Phi)^2]_x.
\end{array}\right.
$}\vskip.05in
This case is the first type of the supersymmetric dark KdV system related to the usual strange dark KdV system. By using the special transformation \eqref{tr1}, (V.a) will be changed to the known strange dark KdV system \eqref{type4} with $a=0$.

\leftline{(VI.a) $ \qquad\qquad
\left\{\begin{array}{l}
\Phi_{t}=-\Phi_{xxx}+6\Phi_x{\cal{D}}\Phi,\\
\Psi_t=-\Psi_{xxx}+3{\cal{D}}[({\cal{D}}\Phi){\cal{D}}\Psi],
\end{array}\right.
$}\vskip.05in
\leftline{(VI.b) $ \qquad\qquad
\left\{\begin{array}{l}
\Phi_{\tau}=-\Phi_{xxxxx}+10(\Phi_{xx}{\cal{D}}\Phi)_x+10\Phi_{x}{\cal{D}}\Phi_{xx}
-30\Phi_x({\cal{D}}\Phi)^2,\\
\Psi_{\tau}=[-\Psi_{xxxx}+(\Phi_{x}{\cal{D}}\Psi)_x+5\Psi_{xx}{\cal{D}}\Phi]_x
+5{\cal{D}}[\Phi_{xx}\Psi_x-({\cal{D}}\Phi)^2{\cal{D}}\Psi].
\end{array}\right.
$}\vskip.05in
Eq. (VI.a) can be viewed as the second supersymmetric dark KdV extension of the strange dark KdV system because \eqref{type4} with $a=0$ can be retrieved from (VI.a) by using the transformation \eqref{tr2}.
\vskip.05in

\leftline{(VII.a) $ \qquad\qquad
\left\{\begin{array}{l}
\Phi_{t}=-\Phi_{xxx}+6\Phi_x{\cal{D}}\Phi,\\
\Psi_t=-\Psi_{xxx}+3\Psi_x{\cal{D}}\Phi,
\end{array}\right.
$}\vskip.05in
\leftline{(VII.b) $ \qquad\qquad
\left\{\begin{array}{l}
\Phi_{\tau}=-\Phi_{xxxxx}+10(\Phi_{xx}{\cal{D}}\Phi)_x+10\Phi_{x}{\cal{D}}\Phi_{xx}
-30\Phi_x({\cal{D}}\Phi)^2,\\
\Psi_{\tau}=-\Psi_{xxxxx}+5(\Psi_{xx}{\cal{D}}\Phi)_x+5[{\cal{D}}\Phi_{xx}-2({\cal{D}}\Phi)^2]\Psi_x.
\end{array}\right.
$}\vskip.05in
The supersymmetric dark KdV system (VII.a) will be reduced back to the usual dual strange dark KdV system \eqref{type5} with $a=0$ through the transformation \eqref{tr1}. \vskip.05in

\leftline{(VIII.a) $ \qquad\qquad
\left\{\begin{array}{l}
\Phi_{t}=-\Phi_{xxx}+6\Phi_x{\cal{D}}\Phi,\\
\Psi_t=(2\Psi_{xx}-6\Psi{\cal{D}}\Phi)_x,
\end{array}\right.
$}
\vskip.05in
\leftline{(VIII.b) $ \qquad\qquad
\left\{\begin{array}{l}
\Phi_{\tau}=-\Phi_{xxxxx}+10(\Phi_{xx}{\cal{D}}\Phi)_x+10\Phi_{x}{\cal{D}}\Phi_{xx}
-30\Phi_x({\cal{D}}\Phi)^2,\\
\Psi_{\tau}=[4\Psi_{xxxx}-20(\Psi_{x}{\cal{D}}\Phi)_x-10\Psi{\cal{D}}\Phi_{xx}+10\Psi({\cal{D}}\Phi)^2]_x.
\end{array}\right.
$}
\vskip.05in
This system corresponds to the first type of suppersymmetric case of the mystery dark KdV system \eqref{type6} with $a=0$. In fact, the transformation \eqref{tr1} can lead Eq. (VIII.a) to \eqref{type6} for $a=0$.

\vskip.05in
\leftline{(IX.a) $ \qquad\qquad
\left\{\begin{array}{l}
\Phi_{t}=-\Phi_{xxx}+6\Phi_x{\cal{D}}\Phi,\\
\Psi_t=2\Psi_{xxx}-6{\cal{D}}[({\cal{D}}\Phi){\cal{D}}\Psi],
\end{array}\right.
$}
\vskip.05in
\leftline{(IX.b) $ \qquad\qquad
\left\{\begin{array}{l}
\Phi_{\tau}=-\Phi_{xxxxx}+10(\Phi_{xx}{\cal{D}}\Phi)_x+10\Phi_{x}{\cal{D}}\Phi_{xx}
-30\Phi_x({\cal{D}}\Phi)^2,\\
\Psi_{\tau}=[4\Psi_{xxxx}-20(\Psi_{x}{\cal{D}}\Phi)_x-10\Phi_{xx}{\cal{D}}\Psi]_x
+10{\cal{D}}[({\cal{D}}\Psi)({\cal{D}}\Phi)^2-\Phi_{xx}\Psi_x].
\end{array}\right.
$}
\vskip.05in
This case can be considered as the second type of suppersymmetric form of the mystery dark KdV system \eqref{type6}. If we apply the transformation \eqref{tr2}, then Eq. (IX.a) becomes \eqref{type6} with $a=0$.
\vskip.05in

\leftline{(X.a) $ \qquad\qquad
\left\{\begin{array}{l}
\Phi_{t}=-\Phi_{xxx}+6\Phi_x{\cal{D}}\Phi,\\
\Psi_t=2\Psi_{xxx}-6({\cal{D}}\Phi)\Psi_x,
\end{array}\right.
$}\vskip.05in
\leftline{(X.b) $ \qquad\qquad
\left\{\begin{array}{l}
\Phi_{\tau}=-\Phi_{xxxxx}+10(\Phi_{xx}{\cal{D}}\Phi)_x+10\Phi_{x}{\cal{D}}\Phi_{xx}
-30\Phi_x({\cal{D}}\Phi)^2,\\
\Psi_{\tau}=4\Psi_{xxxxx}-20(\Psi_{xx}{\cal{D}}\Phi)_x-10\Psi_x[{\cal{D}}\Phi_{xx}-({\cal{D}}\Phi)^2].
\end{array}\right.
$}\vskip.05in
The usual dual mystery dark KdV system \eqref{type7} can be obtained from the supersymmetric dark KdV system (X.a) with $a=b=0$ by utilizing the transformation \eqref{tr1}. \vskip.05in

\leftline{(XI.a) $ \qquad\qquad
\left\{\begin{array}{l}
\Phi_{t}=-\Phi_{xxx}+6\Phi_x{\cal{D}}\Phi,\\
\Psi_t=a\Psi{\cal{D}}\Psi_x+2\Psi_x{\cal{D}}\Phi,
\end{array}\right.
$}\vskip.05in
\leftline{(XI.b) $ \qquad\qquad
\left\{\begin{array}{l}
\Phi_{\tau}=-\Phi_{xxxxx}+10(\Phi_{xx}{\cal{D}}\Phi)_x+10\Phi_{x}{\cal{D}}\Phi_{xx}
-30\Phi_x({\cal{D}}\Phi)^2,\\
\Psi_{\tau}=2[{\cal{D}}\Phi_{xx}-3({\cal{D}}\Phi)^2]\Psi_x
+a[{\cal{D}}\Phi_{xxx}-6({\cal{D}}\Phi){\cal{D}}\Phi_{x}]\Psi.
\end{array}\right.
$}\vskip.05in
The usual ninth type dark KdV system \eqref{type9} with \eqref{abc1} and $b=0$ is a special case of the supersymmetric dark KdV system (XI.a) via the transformation \eqref{tr1}.

\leftline{(XII.a)
$\qquad \qquad
\left\{\begin{array}{l}
\Phi_{t}=-\Phi_{xxx}+6\Phi_x{\cal{D}}\Phi,\\
\Psi_t=a\Psi_{xxx},
\end{array}\right.
$}\vskip.05in
\leftline{(XII.b)
$\qquad \qquad
\left\{\begin{array}{l}
\Phi_{\tau}=-\Phi_{xxxxx}+10(\Phi_{xx}{\cal{D}}\Phi)_x+10\Phi_{x}{\cal{D}}\Phi_{xx}
-30\Phi_x({\cal{D}}\Phi)^2,\\
\Psi_{\tau}=b\Psi_{xxxxx}.
\end{array}\right.
$}\vskip.05in

This supersymmetric dark KdV extension exhibits a trivial and fully decoupled nature, sharing similarities with the standard decoupled dark KdV system described by equation \eqref{type8} when $a$ and $b$ are both set to zero.

Within this subsection, it has been observed that every homogeneous dark KdV system documented in Section 2 possesses, at minimum, one supersymmetric extension. Notably, for the dark KdV systems pertaining to the symmetry case denoted by equation \eqref{type2}, the strange case represented by equation \eqref{type4}, and the mysterious case characterized by equation \eqref{type6}, two distinct categories of supersymmetric dark KdV extensions exist.

\section{Concluding remarks}

In summary, a plethora of dark equation systems can be derived through a variety of methodologies, encompassing the bosonization technique applied to supersymmetric integrable systems and super-integrable models, the dark parameter expansion method, and the decomposition method of higher-dimensional integrable systems. These methodologies offer powerful avenues for investigating the intricate nature of dark equations, unveiling their properties. The existence of an infinite number of such systems underscores the profound complexity and richness inherent in the realm of dark equations, fostering continuous exploration and discovery in this cutting-edge field. Such advancements hold great promise for advancing our understanding of fundamental physics and may have far-reaching implications in various scientific disciplines.
	\\
	\indent The dark equations elucidated in this paper exhibit remarkable integrability properties, primarily due to the presence of higher order symmetries. Indeed, the systems under investigation may exhibit additional integrable properties. These properties encompass a wide range of integrability aspects, including Lax integrability, recursion operators, bi-Hamiltonian structures, B\"acklund and Darboux transformations, conservation laws, and the Painlev\'e property, among others. For instance, the dark KdV system, as represented by equation \eqref{type8}, possesses a recursion operator
$$\phi=\left(\begin{array}{cc} \phi_1 & 0\\ \phi_2 & \gamma_1\partial_x^2
\end{array}\right) $$
with $\phi_i=c_i\partial_x^2-4u-2u_x\partial_x^{-1},\ i=1,\ 2,\ c_1=1,\ c_2=1-\gamma_1$.

Under the framework of Lax pairs and bi-Hamiltonian integrability \cite{Kuper}, the Lax pair ($\hat{\cal{L}} \equiv -4\partial_x^3+6\Phi{\cal{D}}^3+3\Phi_x{\cal{D}}$)
\begin{eqnarray}
&& \left(\begin{array}{cc}\partial_x^2-\Phi{\cal{D}} & 0\\ -\Phi_1{\cal{D}} - q & \partial_x^2-\Phi{\cal{D}}
\end{array} \right)\left(\begin{array}{c}\Psi_1\\ \Psi_2
\end{array} \right)=\left(\begin{array}{cc}\lambda & 0\\ 0 &\lambda
\end{array} \right)\left(\begin{array}{c}\Psi_1\\ \Psi_2
\end{array} \right),\nonumber\\
&& \left(\begin{array}{c}\Psi_1\\ \Psi_2
\end{array} \right)_t=\left(\begin{array}{cc}\hat{\cal{L}} & 0\\ 6\Phi_1{\cal{D}}^3+6r{\cal{D}}^2+3\Phi_{1x}{\cal{D}}+3s & \hat{\cal{L}} -3v
\end{array} \right)\left(\begin{array}{c}\Psi_1\\ \Psi_2
\end{array} \right), \label{SLt}
\end{eqnarray}
defines a supersymmetric Lax integrable dark KdV system
\begin{eqnarray}
&& \Phi_t=(3\Phi{\cal{D}}\Phi-\Phi_{xx})_x,\nonumber\\
&& \Phi_{1t}=(3\Phi{\cal{D}}\Phi_1+\Phi_1{\cal{D}}\Phi+6r \Phi-\Phi_{1xx})_x-3v\Phi_1\nonumber\\
&& q_t=\left[(3\Phi_1\Phi)_x+3\Phi{\cal{D}}r-q_{xx}\right]_x+3\Phi{\cal{D}}q_x-3vq-3\Phi{\cal{D}}s,
\end{eqnarray}
where $\Phi$ and $\Phi_1$ are two fermionic super-fields and $q$ is a bosonic super-field while the other three bosonic potential fields $\{v,\ r,\ s\}$ are related to $\Phi,\ \Phi_1$ and $q$ by
$$v_x=\Phi\Phi_x,\ r_x=q_x+\Phi\Phi_1,\ s_x=(\Phi_1\Phi+q_x)_x-\Phi{\cal{D}}(q-r).$$
\indent
According to the findings presented in this paper, the notion of dark equations extends beyond a mere conceptual analogy to dark matter and dark energy. In fact, there exists a tangible connection between dark equations and dark matter, as various dark equations can be directly associated with supersymmetric systems. These systems, in turn, are intricately linked to supergravity and dark matter phenomena. The results of this study indicate the presence of multiple dark fields within a single supersymmetric framework. This revelation underscores the complexity and richness of dark equations within the context of supersymmetry, shedding light on the potential interplay between these equations and the enigmatic nature of dark matter.
The presence of such diverse integrable features highlights the profound mathematical structure inherent in these systems. The identification and exploration of these properties pave the way for a deeper understanding of the intricate dynamics and profound connections associated with dark equations, unraveling the intricate interplay between various integrable structures and shedding light on the fundamental nature of these systems.

\section*{Acknowledgments}
The work was sponsored by the National Natural Science Foundations of China (Nos.12235007, 11975131). The author is indebt to thank Profs. X. Z. Hao, Q. P. Liu, R. X. Yao, M. Jia, X. B. Hu and K. Tian for their helpful discussions.

\end{document}